\documentclass[ba]{imsart}
\pubyear{0000}
\volume{00}
\issue{0}
\doi{0000}
\firstpage{1}
\lastpage{1}

\usepackage{amsthm}
\usepackage{amsmath}
\usepackage{natbib}
\usepackage[colorlinks,citecolor=blue,urlcolor=blue,filecolor=blue,backref=page]{hyperref}
\usepackage{graphicx}
\usepackage[english]{babel}
\usepackage[utf8]{inputenc}
\usepackage{amsmath}
\usepackage{amssymb}
\usepackage{graphicx}
\usepackage{hyperref}
\usepackage{graphicx}
\usepackage{comment}
\usepackage{mathrsfs}
\usepackage{amsthm}
\usepackage{bbm}
\usepackage{rotating}
\usepackage{lscape}
\usepackage[colorinlistoftodos]{todonotes}
\usepackage{lipsum}
\usepackage{comment}
\usepackage{natbib}
\usepackage[colorlinks,citecolor=blue,urlcolor=blue,filecolor=blue,backref=page]{hyperref}
\usepackage{graphicx}
\newtheorem{theorem}{Theorem}

\startlocaldefs
\numberwithin{equation}{section}
\theoremstyle{plain}

\endlocaldefs

\begin{document}
\nocite{multiMVT}
\nocite{MCMCboot}
\begin{frontmatter}
\title{High-dimensional single-index Bayesian modeling of brain atrophy}
\runtitle{Bayesian single-index model for the Brain Atrophy}

\begin{aug}
\author{\fnms{Arkaprava} \snm{Roy}\thanksref{addr2,t1}\ead[label=e1]{aroy2@ncsu.edu}},
\author{\fnms{Subhashis} \snm{Ghosal}\thanksref{addr1,t1}\ead[label=e2]{sghoshal@ncsu.edu}}
\author{\fnms{Kingshuk} \snm{Roy Choudhury}\thanksref{addr3}
\ead[label=e3]{kingshuk.roy.choudhury@duke.edu}}
\and
\author{\fnms{For The Alzheimer's Disease Neuroimaging Initiative}\thanksref{t2}}
\runauthor{Roy et al.}

\address[addr2]{Department of statistics,
Duke University,
    \printead{e1} 
}

\address[addr1]{Department of statistics,
North Carolina State University, 
    \printead*{e2}
}

\address[addr3]{Department of Biostatistics and Bioinformatics, Duke University Medical Center, Durham, NC, \printead{e3}
}

\thankstext{t1}{Research is partially supported by NSF grant DMS-1510238.}
\thankstext{t2}{Data used in the preparation of this article were obtained from the Alzheimer's Disease Neuroimaging Initiative
(ADNI) database (adni.loni.usc.edu). As such, the investigators within the ADNI contributed to the design
and implementation of ADNI and/or provided data but did not participate in analysis or writing of this report.
A complete listing of ADNI investigators can be found at:
\url{http://adni.loni.usc.edu/wp-content/uploads/how_to_apply/ADNI_Acknowledgement_List.pdf}}

\end{aug}

\begin{abstract}
We propose a model of brain atrophy as a function of high-dimensional genetic information and low dimensional covariates such as gender, age, APOE gene, and disease status. A nonparametric single-index Bayesian model of high dimension is proposed to model the relationship with B-spline series prior on the unknown functions and Dirichlet process scale mixture of centered normal prior on the distributions of the random effects. The posterior rate of contraction without the random effect is established for a fixed number of regions and time points with increasing sample size. We implement an efficient computation algorithm through a Hamiltonian Monte Carlo (HMC) algorithm. The performance of the proposed Bayesian method is compared with the corresponding least square estimator in the linear model with horseshoe prior, LASSO and SCAD penalization on the high-dimensional covariates. The proposed Bayesian method is applied to a dataset on volumes of brain regions recorded over multiple visits of 748 individuals using 620,901 SNPs and 6 other covariates for each individual, to identify factors associated with brain atrophy. 
\end{abstract}


\end{frontmatter}

\section{Introduction}
\label{introduction}

Alzheimer's disease (AD) is a progressive neurodegenerative disease that affects approximately 5.5 million people in the United States and about 30 million people worldwide. It is believed to have a prolonged preclinical phase initially characterized by the development of silent pathologic changes when patients appear to be clinically normal, followed by mild cognitive impairment (MCI) and then dementia (AD) (\cite{Petrella}). Apart from its manifestation in the impairment of cognitive abilities, disease progression also produces a number of structural changes in the human brain, which includes the deposition of amyloid protein and the shrinkage or atrophy for certain regions of the brain over time (\cite{Thompson}). Previous studies have shown that the rate of brain atrophy is significantly modulated by a number of factors, such as gender, age, baseline cognitive status and most markedly, allelic variants in the Apolipoprotein E (APOE) gene (\cite{Hostage}). In this paper, we examine if any other genes are also implicated in modulating the rate of brain atrophy along with examining effects of the low dimensional covariate on the rate of atrophy using the data, collected by Alzheimer's Disease Neuroimaging Initiative (ADNI). 



\begin{figure}[htbp]
\centering
\includegraphics[width = .5\textwidth]{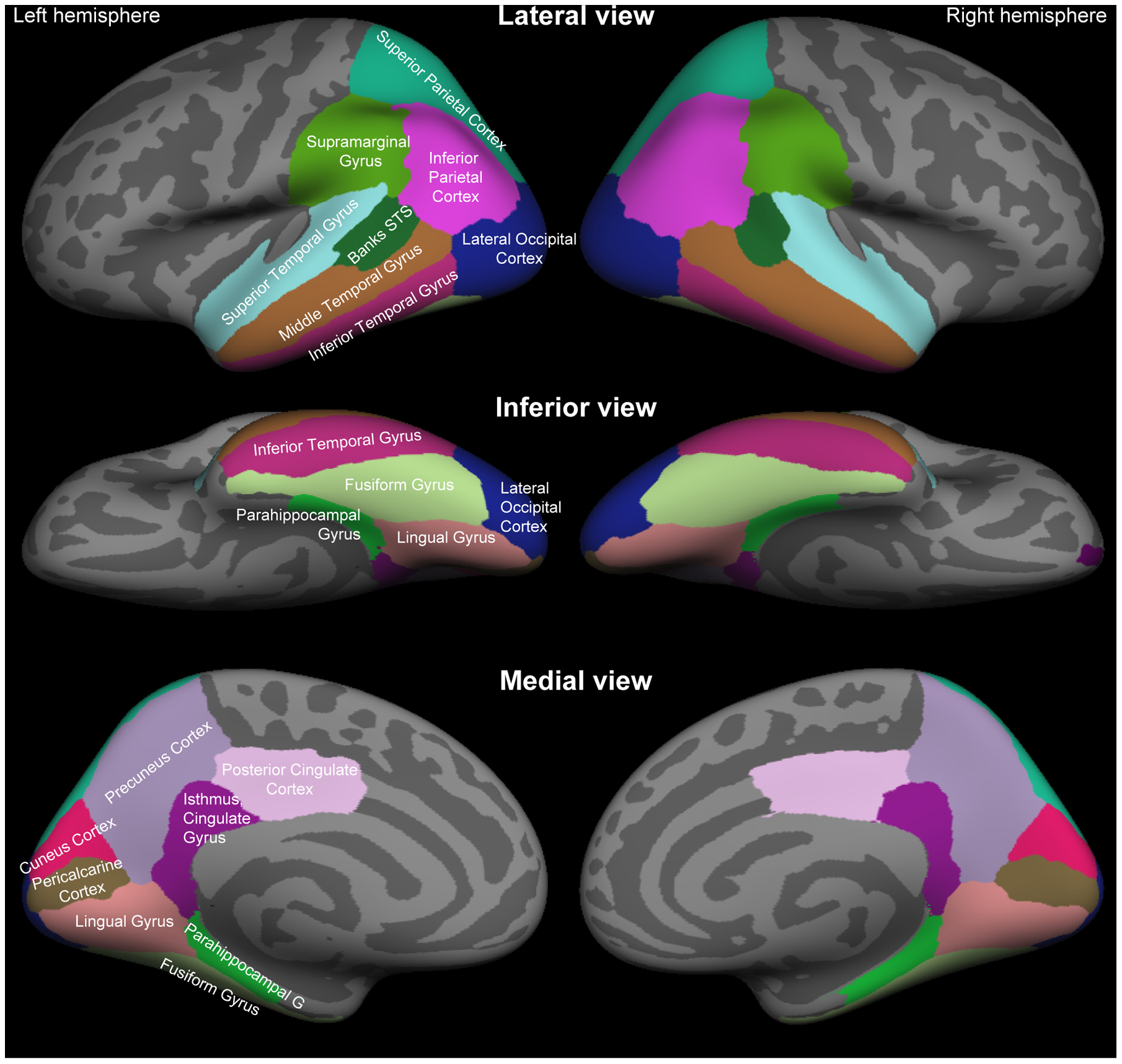}
\caption{Anatomic parcellation of cortical surface from different angles showing brain regions used for analysis. (courtesy: \cite{brain})}
\label{brain}
\end{figure}

We model regional brain volume, which has been measured from magnetic resonance (MR) images using a segmentation procedure and recorded in the ADNI database. We collect this data directly from ADNI. We have volumetric measurements over six visits of thirteen disjoint brain regions and a total brain measure which is a summary measure of total brain parenchyma, including the Cerebral-Cortex, Cerebellum-Cortex, Thalamus-Proper, CaudatePutamen, Pallidum, Hippocampus, Amygdala, Accumbens-area, VentralDC, Cerebral-White-Matter, Cerebellum-White-Matter, and WM-hypointensities. These visits are roughly around six months apart. Thus the subjects are scanned roughly for three years. The anatomic structures of the brain also differ across different individuals and are assumed to be dependent on subject-specific covariates like genetic variations, gender, age etc. In this paper, we propose a model to study the effects of these different covariates along with Alzheimer's disease state on brain atrophy in different brain regions. This analysis represents a technical challenge because the genomic data is high-dimensional and needs to be incorporated in a model for longitudinal progression of brain volumes measured in multiple parts of the brain in a non-parametric setup. A schematic of the regions, we studied here, are depicted in the Figure~\ref{brain}. This image is obtained from \cite{brain}. Although our analysis does not have any association with this paper, an image from it is used to show the brain regions used for the analysis in our paper. We have not collected the data directly from the magnetic resonance (MR) images according to some brain parcellation atlas. We got the data in a preprocessed form from ADNI itself. \citep{Hostage} had also used a similar set of regions of interest. 

We consider two separate sets of unknown functions of covariates $X$ and $Z$ to model the volumetric measurements of the first visit and rates of changes for different regions. These functions have two inputs. The first $X$ consists of high-dimensional single-nucleotide polymorphism (SNP) of each individual, and the other $Z$ consists of low-dimensional covariates like gender, age, disease state, APOE gene status of each individual. The effect of these covariates on brain volume is modelled using a semi-parametric function of the form  \{$a_{0,j}(X'\beta,Z'\eta): j=1,\ldots,14$\} for the volumetric measurements of the first visit and \{$a_{1,j}(X'\beta,Z'\eta): {j=1,\ldots, 14}$\} for the rate of change in the $j$th region and the coefficients $\beta$ and $\eta$ are unit vectors of appropriate dimensions. A finite random series prior is put on the functions based on tensor products of B-splines with appropriate prior distribution on the coefficients. To take care of the high-dimensionality of $X$, the coefficient $\beta$ is assumed to be sparse. We reparametrize $\beta$ in polar coordinates to incorporate sparsity in its prior. We incorporate the effect of time in the modeling by an increasing function which is estimated nonparametrically also using a finite random series of B-splines with an appropriate prior on the coefficients.

Apart from proposing a sophisticated model for studying brain atrophy, the proposed method develops a new estimation scheme for a general high-dimensional single-index model. Estimation for high-dimensional single-index model was previously addressed in \cite{Zhu&Zhu}, \cite{Yu&Ruppert}, \cite{Wang}, \cite{Peng&Huang}, \cite{Radchenko} and \cite{Luo&Ghosal}. All of them used the $\ell_1$-penalty and used optimization techniques to get the estimates. In a Bayesian framework, \cite{Antoniadis} used the Fisher–von Mises prior for the directional vector. This cannot be easily modified for a high-dimensional covariate as then we shall need a prior which favors many zeros in the unit vector. Another paper addressing sparse Bayesian single-index model estimation is \cite{Alquier}. Even though their method is theoretically attractive, it is difficult to implement for high-dimensional covariate due to its high computational complexity. \cite{Wangh} developed a Bayesian method for sparse single-index model using the reversible jump Markov chain Monte Carlo (MCMC) technique which is computationally expensive, especially in the high-dimensional setting. We introduce a new way of incorporating sparsity on a unit vector by a spike and slab prior via the polar form. The computation scheme is based on an efficient gradient based Hamiltonian Monte Carlo (HMC) algorithm. 

The rest of the paper is organized as follows. Section~\ref{data description} discusses the dataset and modeling in more detail. In Section~\ref{prior}, we describe the prior on different parameters of the model. Section~\ref{computation} describes posterior computation in this setup. We study the posterior rate of contraction in the model in Section~\ref{consistency} under the asymptotic regime that the number of individuals goes to infinity but the number of time points where measurements are taken and the number of regions is fixed. We present a simulation study comparing the proposed Bayesian procedure with its linear counterpart in Section~\ref{simulation}. The concentration of the posterior justifies the use of the proposed Bayesian procedure from a frequentist perspective. In Section~\ref{realdata}, we present conclusions from the ADNI data on brain atrophy described above using our proposed method. Section~\ref{conclusions-discussion}, concludes the paper with some further remarks. 

\section{Data description and modeling}
\label{data description}
Data used in the preparation of this article were obtained from the Alzheimer's Disease Neuroimaging Initiative (ADNI) database (\url{adni.loni.usc.edu}). ADNI was launched in 2003 as a public-private partnership, led by Principal Investigator Michael W. Weiner, MD. The primary goal of ADNI has been to test whether serial magnetic resonance imaging (MRI), positron emission tomography (PET), other
biological markers, and clinical and neuropsychological assessment can be combined to predict the progression of mild cognitive impairment (MCI) and early Alzheimer's disease (AD). In the ADNI dataset, the grey matter part of the brain is divided into thirteen disjoint regions. The volume of these regions and the summary measure of the whole brain are recorded over time for $n=748$ individuals. The visits are scheduled after each 6 months over a span of 36 months. Not all of the participants turned up at each of those scheduled visits. There was no record of visiting after 30-$th$ month for any of the participants. The volume data of $J=14$ regions which include thirteen brain regions and the summary measure of the whole brain over $T_i$ set of time points for the $i$th individual, for $i=1,\ldots ,748$ is collected where $1 \leq |T_i|\leq 6$. The notation $|T_i|$ denotes cardinality of the set $T_i$. For some of the participants, the disease status changed during the span of this study. They were very small in number. We do not include them in this study. 

In ADNI, the subjects were genotyped using the Human 610-Quad BeadChip (Illumina, Inc., San Diego, CA), yielding a set of 620,901 SNP and copy number variation (CNV) markers. The APOE gene has been the most significant gene in GWAS of Alzheimer's disease. The corresponding SNPs, rs429358 and rs7412, are not on the Human 610-Quad Bead-Chip. At the time of participant enrollment, APOE genotyping was performed and included in the ADNI database. The two SNPs (rs429358, rs7412) define the epsilon 2, 3, and 4 alleles and therefore were not genotyped using DNA extracted by Cogenics from a 3 mL aliquot of EDTA blood. These alleles are considered separately in the study.

Thus apart from the volumetric measurements, we also have high-dimensional SNP data and data on some other covariates for each individual. The other covariates are gender, disease state, age and allele 2 and 4 of the APOE gene. Except for the covariate age, all other low-dimensional covariates are categorical. To represent the categories, we use binary dummy variables. Since the disease status has three states--- NC (no cognitive impairment), MCI (mild cognitive impairment) and AD (Alzheimer's disease), we consider two dummy variables $Z^{\text{AD}}$ and $Z^{\text{MCI}}$ respectively standing for the onset of MCI and AD, setting NC at the baseline. Similarly, the dummy variable $Z^{\mathrm{M}}$ indicating male gender is introduced setting females at the baseline. Also, we introduce $Z^{\text{APOE},2}$, $Z^{\text{APOE},4}$ standing for Alleles 2 and 4 for the two alleles APOEallele2 and APOEallele4 together setting Allele 3 as a baseline for each of the two cases. We consider the age corresponding to the initial visit as a covariate as well. Let $Z=(Z^{\mathrm{MCI}},Z^{\mathrm{AD}},Z^{\mathrm{M}},\mathrm{Age}, Z^{\text{APOE},2}, Z^{\text{APOE},4})$ stand for the whole vector of covariates. The continuous variable, Age, is standardized.

With time, different brain regions change differently. We study the effects of different attributes to these changes. For every individual, the volume of a brain region on a particular visit should primarily depend on the volume of that region at the zeroth visit and the rate of change of volume for that region with time. These rates of changes are region-specific as well as individuals-specific. Hence, it is logical to consider the baseline volume and the rate of change as functions on the subject and brain region. As the geometry of brain structure is complicated, we do not assume a form of standard spatial dependence between measurements across brain regions. Thus we need two sets \{$a_{0,j}(\cdot, \cdot),a_{1,j}(\cdot, \cdot): {j=1,\ldots, 14}$\} of functions for modeling volume at the initial visit and the rate of change for the $j$th region. These functions are unknown and are modeled nonparametrically. For nonparametric regression problems, single-index models provide a lot of flexibility in estimation and interpretation of the results. Hence we adopt the bivariate single-index model with two inputs for high-dimensional and low-dimensional covariates separately for easy interpretation and computational efficiency. The effect of time is captured through an unknown increasing function $F_0(\cdot)$, which is bounded in $[0,1]$. This is also modeled nonparametrically. 

Let $Y_{ijt}$ be the volume of the $j$th brain region for the $i$th individual at the $t$th time point in the logarithmic scale, $X_{i}$ is high-dimensional SNP expression of length $p$ for the $i$th individual, $t\in T_i/36$, $i=1,\ldots,m$ and $j=1,\ldots,14$. The visit times are divided by 36 to keep it bounded in [0, 1]. Then 
the data generating process can be represented through the following specification
\begin{gather}
Y_{ijt} = F_{ijt} + \varepsilon_{ijt}, \varepsilon_{ijt} \sim^{\mathrm{iid}} \mathrm{N}(0, \sigma^2) \label{model}, \\
F_{ijt} = a_{0,j}(X_i'\beta, Z_{i}'\eta) - a_{1,j}(X_i'\beta, Z_{i}'\eta)F_0(t), \nonumber
\end{gather}
where $a_{0,j}(\cdot, \cdot), a_{1,j}(\cdot, \cdot), F_0(\cdot)$ are all unknown continuous functions and N stands for a normal distribution. The function $F_0(\cdot)$ is monotone increasing functions from $[0, 1]$ to $[0, 1]$ and $F_0(0)=0$, $F_0(1)=1$. The other two functions $a_{0,j}(\cdot, \cdot), a_{1,j}(\cdot, \cdot)$ maps from $[-1,1]^2$ to $(-\infty, \infty)$. For identifiability of the functions along with the parameters $\beta$ and $\eta$, we assume that $\|\beta\| = 1$, $\|\eta\|=1$; here $\|\cdot\|$ denotes $L_2$-norm of a vector. We also normalize the covariates for each individual i.e. $X_i$ and $Z_{i}$ for each $i$ such that $\|X_i\|$ and $\|Z_i\|$ are one. This is to ensure that $X_i'\beta$ and $Z_{i}'\eta$ are bounded between $[-1, 1]$ for each $i$. For nonparametric function estimation, bounded domain is important for uniform approximation using the basis expansion. The biggest challenge for  estimation in this model is the high-dimensionality of $\beta$. To identify important SNPs from $X$, we need to perform variable selection. To do that, we propose a sparse estimation scheme for $\beta$. First we reparametrize the two unit vector $\beta=(\beta_1,\ldots,\beta_{p})$ and $\eta=(\eta_1,\ldots,\eta_{k})$ to their respective polar forms which allow us to work with Euclidean spaces. In this setup, only the directions of $\beta$ and $\eta$ are identifiable. Note that $\beta$ and $-\beta$ have the same directions. In the polar setup, for $s =1,\ldots, p-1$, $\beta_s = \prod_{l=1}^{s-1}\sin\theta_l\cos\theta_s$, and  $\beta_p = \prod_{l=1}^{p-1}\sin\theta_l$ where $\{\theta = (\theta_1,\ldots,\theta_{p-1})\}$ is the polar angle corresponding to the unit vector $\beta$. Here $\theta_s\in [0,\pi]$ for $s=1,\ldots, (p-2)$ and $\theta_{p-1}\in[0,2\pi]$. Similarly, let $\alpha$ be the polar angle corresponding to $\eta$. Then for $s\leq k-1$, $\eta_s = \prod_{l=1}^{s-1}\sin\alpha_l\cos\alpha_s$ and $\eta_k = \prod_{l=1}^{k-1}\sin\alpha_l$.

\section{Prior specification}
\label{prior}
In the nonparametric Bayesian setup described above, we induce prior distributions on the smooth functions $a_{0,j}$ and $a_{1,j}$ in~\eqref{model} through basis expansions in tensor products of B-splines and suitable prior for the corresponding coefficients. Given other parameters in this setup, a normal prior distribution on the coefficients of the tensor products of B-splines will lead to conjugacy and faster sampling via Gibbs sampling scheme. An inverse gamma prior on $\sigma^2$ is an obvious choice due to conjugacy and faster sampling. We also put a B-spline series prior on the smooth increasing function of time $F_0(\cdot)$. The coefficients for this function would be increasing in the index of the basis functions and lie in (0,1]. To put a prior on an increasing sequence, we introduce a set of latent variables of size equal to one less than the number of B-spline coefficients. Then the B-spline coefficients would be normalized cumulative sum of those latent variables. Other two parameters $\beta$ and $\eta$ are reparametrized into their polar coordinate system. The parameter space of the polar angles will be a hyper-rectangle. It will be easier to put prior on the polar angles. To estimate using the sparsity of $\beta$, we need to carefully put a shrinkage prior to the polar angles. A polar angle of $\pi/2$ will ensure that the corresponding coordinate in the unit vector equals to zero. When there is sparsity in the unit vector, most of the polar angles will be $\pi/2$. Thus a spike and slab prior to polar angle with a spike at $\pi/2$ should be able to capture sparsity in the corresponding unit vector. The last polar angle has spike both at the multiples of $\pi$ and $\pi/2$, due to the special structure of the last and the penultimate co-ordinates of a unit vector in the polar form.  If it a multiple of $\pi$, then the last coordinate is zero and if it is an odd multiple of $\pi/2$, then the penultimate coordinate is zero. The support of the last polar angle is $[0, 2\pi]$ and we need spikes at all multiples of $\pi$ and $\pi/2$. The prior is constructed to maintain that structure of spikes. Since only the directions of $\beta$ and $\eta$ are identifiable, the intercept and slope functions are modeled as even functions i.e. symmetric around zero.

Now we describe the prior in details. Let $\lceil x \rceil$ denote the lowest integer greater than or equal to $x$.

\begin{figure}[htbp]
\centering
\includegraphics[width = .88\textwidth]{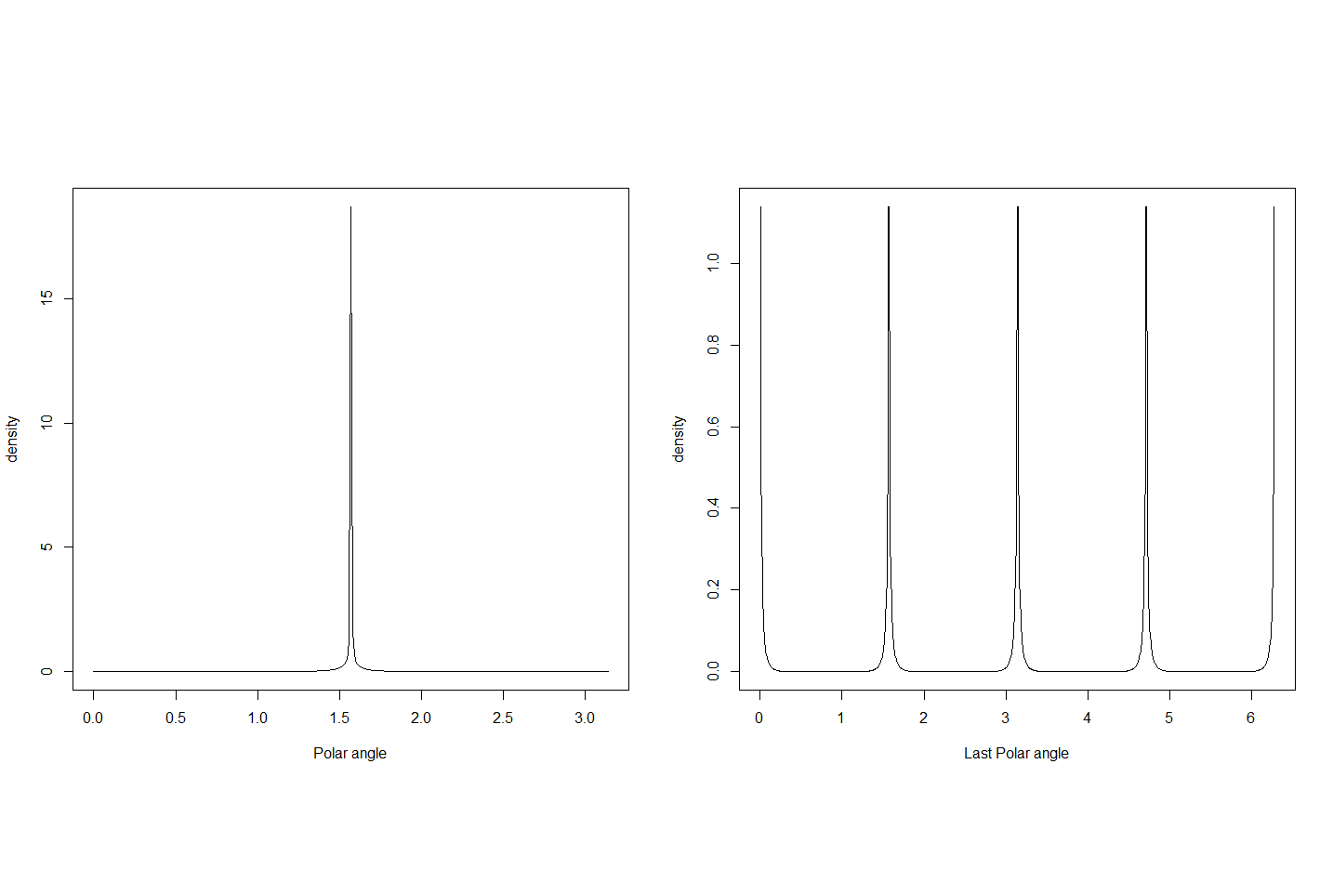}
\caption{Spike distribution with $M_1=0.01$ and $M_2 = 10$.}
\label{priorfig}
\end{figure}

\begin{itemize}
\item[(i)] Intercept and slope functions: Let $a_{\nu,j}(x, y) = \sum_{m=1}^K \sum_{m'=1}^K \lambda_{\nu,j, mm'}B_m(x)B_m'(y)$, $\nu=0,1$, 
with $\lambda_{\nu,j,mm'}=\lambda_{\nu,j,(K-m)m'}$, $\lambda_{\nu,j,mm'}=\lambda_{\nu,j,m(K-m')}$ for $\nu=0,1$, and 
$\lambda_{\nu,j,mm'} \sim \text{N} (0, a^2)$, $1\le m, m'\le \lceil K/2 \rceil$, 
for some chosen $a>0$. Further, $K$ is given a prior with probability mass function of the form $\Pi(K=k)=b_1\exp[-b_2 k^2 (\log k)^{b_3}]$, with $b_1,b_3>0$ and $0\le b_3\le 1$. 

\item[(ii)] The function of time: Let $F_0(x) = \sum_{l=1}^{K'}\lambda_{l}B_l(x)$, 
with $0=\lambda_1\le \lambda_2\le \cdots \le \lambda_{K'} = 1$. We put a prior on $(\lambda_2,\ldots,\lambda_{K'-1})$ by reparameterizing as  
$\lambda_{l+1}=({\sum_{i=1}^{l}\delta_i})/({\sum_{i=1}^{K'-1}\delta_{i}})$, and putting the prior $\delta_{i}\sim \text{Un}(0,1)$ for $l=1,\ldots,K'-1$, where Un stands for the uniform distribution. Further, $K'$ is given a prior with probability mass function of the form $\Pi(K'=k)=b_1'\exp[-b_2' k (\log k)^{b_3'}]$ with $b_1',b_2'>0$ and $0\le b_3'\le 1$. 

\item [(iii)] Error variance: We put $\sigma^{-2}\sim \mathrm{Ga}(d_1,d_2)$, where Ga stands for the gamma distribution. 

\item[(iv)] Polar angles $\alpha$ of $\eta$: We let $\alpha_r\sim \text{Un}(0,\pi)$,  $r=1,\ldots,(k-2)$, and $\alpha_{k-1}\sim \textrm{Un}(0,2\pi)$, independently. 

\item[(v)] Polar angles $\theta$ of $\beta$: We put a spike and slab prior on the polar angles that has spike at $\pi/2$ for the first polar angle and all the multiples of $\pi/2$ for the last polar angle. Then the spike distribution would look like Figure~\ref{priorfig}. The spike and slab prior for $\theta_r$, $r\leq (p-2)$, has density given by 
$$ (1-\gamma_i)\frac{\Gamma(M_1+M_2)}{\Gamma(M_1)\Gamma(M_2)}\bigg(\frac{\min(\theta_r, \pi-\theta_r)}{\pi/2}\bigg)^{M_2}\bigg(1-\frac{\min(\theta_r, \pi-\theta_r)}{\pi/2}\bigg)^{M_1} + \gamma_i \frac{1}{\pi},$$ 
for $0<\theta_r<\pi$ and the distribution of $\theta_{p-1}$ is given by  
\begin{eqnarray*} 
\theta_{p-1} &\sim & (1-\gamma_i)\frac18 [\mathrm{Be}(M_1, M_2)_{[0, \pi/4]}+\mathrm{Be}(M_2, M_1)_{[\pi/4, \pi/2]}\\
&& +\mathrm{Be}(M_1, M_2)_{[\pi/2, 3\pi/4]}+\mathrm{Be}(M_2, M_1)_{[3\pi/4, \pi]}+\mathrm{Be}(M_1, M_2)_{[\pi, 5\pi/4]}\\
&& +\mathrm{Be}(M_2, M_1)_{[5\pi/4, 3\pi/2]}+ \mathrm{Be}(M_1, M_2)_{[3\pi/2, 7\pi/4]}+\mathrm{Be}(M_2, M_1)_{[7\pi/4, 2\pi]}]\\ 
&& +\gamma_i\text{Un}(0,2\pi);
\end{eqnarray*}
here Be$(M_1, M_2)_{[a,b]}$ denotes the beta distribution with shape parameters $M_1$ and $M_2$, supported within the interval $[a,b]$, and $M_1<1\leq M_2$. The indicator variable $\gamma\sim $ Ber$(q)$.
\end{itemize}

The spike distribution on $\theta$ looks like Figure~\ref{priorfig}. The first plot is for the first $(p-2)$ angles and the second plot is for the last polar angle.

Note that either geometric or Poisson distribution (respectively $b_3'=0$ and $1$) can be chosen as prior on $K'$, the number of terms to be used in the B-spline series for the growth function $F_0$. For $K$, the square, which is the number of terms in the tensor product series representation, can have a geometric or Poisson-like tail.

{\it{Model selection}}: Polar angles with a posterior probability of selection in the model more than 0.5 are considered in the model.

\section{Computation}
\label{computation}
Now the conditional log-likelihood is given by 
\begin{align*}
 &{C} - \sum_{ijt} \frac{1}{2\sigma^2}\big(Y_{ijt} - \sum_{m=1}^K \sum_{m'=1}^K \lambda_{0,j, mm'} B_m(\sum_{s=1}^{p-1}X_{is}\prod_{l=1}^{s-1}\sin \theta_l\cos\theta_s +X_{ip}\prod_{l=1}^{p-1}\sin\theta_l)\\
& \qquad \times B_m'(\sum_{s=1}^{k-1}Z_{is}\prod_{l=1}^{s-1}\sin\alpha_l\cos\alpha_s + Z_{ik}\prod_{l=1}^{k-1}\sin\alpha_l)\\
&\quad + \sum_{m=1}^K \sum_{m'=1}^K \lambda_{1,j, mm'} B_m(\sum_{s=1}^{p-1}X_{is}\prod_{l=1}^{s-1}\sin\theta_l\cos\theta_s + X_{ip}\prod_{l=1}^{p-1}\sin\theta_l) \\ 
& \qquad \times B_m'(\sum_{s=1}^{k-1}Z_{is}\prod_{l=1}^{s-1}\sin\alpha_l\cos\alpha_s + Z_{ik}\prod_{l=1}^{k-1}\sin\alpha_l)\sum_{m=1}^{K-1} \frac{\sum_{i=1}^{m}\delta_i}{\sum_{i=1}^{K'-1}\delta_{i}}B_{m+1}(t)\big)^2 \\ 
& \quad - \sum_{m,m',j}\frac{\lambda_{0,j,mm'}^2 + \lambda_{1,j,mm'}^2 }{2a^2} \\ 
&\quad +\sum_{l=1}^{p-1}\log\big((1 - \gamma_l){\theta_l'^{M_1-1}(1-\theta_l')^{M_2-1}} \frac{\Gamma(M_1+M_2)}{\Gamma(M_1)\Gamma(M_2)}+ \gamma_l\big)  \\ &\quad - (J\sum_iT_i/2 + d_1-1) \log \sigma^2 - d_2/\sigma^2,
\end{align*}
where $C$ involves only the hyperparameters $a, M_2,M_1, K,d_1, d_2$ and the observations but not parameter of the model. 

All B-spline coefficients and $\sigma$ are updated using the conjugacy structure. Using derivative of B-splines \citep{de2001practical}, it is possible to calculate the derivatives of the log-likelihood with respect to the polar angles and $\delta$'s of $F_0(t)$. Thus, we implement efficient gradient based MCMC algorithm using Hamiltonian Monte Carlo algorithm, described in \citep{neal2011mcmc}. Since these parameters have bounded support, the candidates for a Metropolis step are reflected back if they cross the boundaries. For example if a candidate $\theta_i^c$ of $\theta_i$ is more than $\pi$, then it is re-adjusted as $\theta_i^c=\pi-(\theta_i^c-\pi)$. Similar adjustment is done if $\theta_i^c$ becomes smaller than zero as well. Similar strategies are adopted while updating $\delta$'s as well. All the posterior updates are discussed in the supplementary materials.

In our computation of the model, we are not using the priors on $K$ and $K'$ i.e. the number of B-spline basis functions as it will require reversible jump MCMC strategy which is computationally expensive. Thus, these are chosen by minimizing the Bayesian Information Criterion (BIC) after fitting the model over a grid of a number of B-spline basis functions for randomly generated 10 different choices of $\beta$ and $\eta$. We generate 10 different choices for $\beta$ and $\eta$ and then fit the non-linear model for different choices of a number of B-spline basis functions within the range [7, 20]. After taking an average overall 10 BIC values for each case, we take the number as optimal that has the least BIC value. Convergence of the MCMC chain is diagnosed using trace plots.  

\section{Large-sample Properties}
\label{consistency}

In this section, we examine large sample properties of the proposed Bayesian procedure for the model \eqref{model}. We have observations for fixed $J$ number of regions and $T$ many time points. We show posterior consistency in the asymptotic regime of increasing sample size and increasing dimension $p$ of SNPs.

We study posterior contraction rate with respect to the empirical $\ell_2$-distance on the regression function, which is defined as follows. For two sets of parameters $(F,a,\beta,\eta)$ and $(F^*,a^*,\beta^*,\eta^*)$, the empirical $\ell_2$-distance is given by  
\begin{eqnarray*}
&& d^2((F,a,\beta,\eta),(F^*,a^*,\beta^*,\eta^*)) \\
&& \quad = \frac{1}{J \sum_{i=1}^n T_i} \sum_{j=1}^J \sum_{i=1}^n \sum_{t=1}^{T_i} |a_{0,j}(X_i'\beta, Z_i'\eta)+a_{1,j}(X_i'\beta,Z_i'\eta) F(t)\\
&& \qquad \qquad \qquad \qquad \qquad \qquad -a^*_{0,j}(X_i'\beta^*, Z_i"\eta^*)-a^*_{1,j}(X_i'\beta^*,Z_i'\eta^*)F^*(t) |^2,
\end{eqnarray*}
and $a=(a_{0,j}, a_{1,j}: j=1,\ldots,J)$, $a^*=(a_{0,j}^*, a_{1,j}^*: j=1,\ldots,J)$.

Since $\beta$ is a high-dimensional parameter, sensible estimation is possible only if it has sparsity, which must be picked up by the prior. In the setting of a spike-and-slab prior for polar coordinates described in Section~\ref{prior}, we need to ensure sufficient concentration near $\pi/2$ by choosing a large value of the second parameter $M_2$ in the beta spike distribution (depending on the sample size $n$ and the dimension $p$) and a small value of the probability of slab $\gamma$. More precisely, we choose $M_1\le 1$ fixed, $M_2>\sqrt{np}\log p$ and $\gamma=o(p^{-1})$. Then the  contraction rate will be determined by the smoothnesses of the underlying true functions $a_{0,0},a_{0,1}$ and $F_{0,0}$, the sparsity $s_0$ of true high-dimensional regression coefficient $\beta_0$ and mildly on the parameter $b_3,b_3'$ in the prior distribution for the number of basis elements $K,K'$ used in the B-spline bases, as shown by the following result. 

\begin{theorem}
Assume that the true function $F_{0,0}$ belongs to the H\"older class of regularity level $\iota'$ on $[0,1]$ and the true coefficient functions $(a_{0,0,j},a_{1,0,j}:, j=1,\ldots,J)$ are H\"older smooth function of regularity level $\iota$ on $[-1,1]\times [-1,1]$. Let the true regression coefficient $\beta_0$ for the high-dimensional covariate $X$ have $s_0$ non-zero co-ordinates. Then the posterior contraction rate with respect to the distance $d$ is $$
\max\bigg\{n^{-\iota/(2\iota+2)} (\log n)^{\iota/(2\iota+2)+(1-b_3)/2},n^{-\iota'/(2\iota'+1)} (\log n)^{\iota'/(2\iota'+1)+1-b_3'},\sqrt{\frac{s_0\log p}{n}}\bigg\}. 
$$
\end{theorem}

In the above result, since the observation points are not dense over the domain, posterior contraction on the function is based on its distance with the true function only at the observation points. 

The proof of the theorem uses the general theory of posterior contraction (see \cite{Ghosal}) for independent non-identically distributed observations and some estimates for finite random series based on B-splines. The proof is given in the supplementary materials part.

\section{Simulation}
\label{simulation}
We compare our method with some common penalization methods based on the following simplified linear model 
\begin{align}
\log Y_{ijt} = X_i'\beta + Z_i'\eta + \gamma_{1,j} - (X_i'\beta + Z_i'\eta + \gamma_{2,j})t +\varepsilon_{ijt}, \label{linear}
\end{align}
$t=1,\ldots,T_i$, $i=1,\ldots,n$, $j=1,\ldots,13$.
In the above model, $\beta$ is a sparse vector and all other parameters are unpenalized. The performance of these methods is compared based on MSE values on a test set under the scenarios the linear model is correct and the linear model is false. 

We generate a high dimensional binary matrix $X$ and a low-dimensional covariate matrix $Z$. The true data matrices of the real data are in the support of the scheme of sampling the high dimensional matrix as well as the low dimensional matrix.

{\it{Data generation for the non-linear case:}}

\begin{itemize}
\item Generate a data matrix $X$ with elements coming from Bernoulli distribution with success probability of the $i$th row as $p_i$. 

\item Generate $p_i$, $i=1,\ldots,n$, from the standard uniform distribution.

\item Generate all the elements of the matrix $Z$ from N$(0,1)$.

\item Generate the sparse vector $\beta_0$ with 5\% elements non-zero. Positions for non-zero elements are chosen first at random by sampling $p/20$ elements from total $p$ positions, where $p$ is the length of $\beta_0$. The non zero elements are generated from mixture distribution of two normals N$(2,1)$ and N$(-1,1)$.

\item Set the value of $\eta_0$ to $(1, - 2, 4.3, 10, -8)$.

\item Normalize each row of $X$ and $Z$ along with $\beta_0$ are $\eta_0$ to the unit norm.
\end{itemize}

We let the true functions be $$a_{0,0,j}(x, y) = 2((j/13)x)^3+2((1-j/13)y)^3,$$ 
$$a_{0,1,j}(x, y) = 2\exp((j/13)y)+2\exp((1-j/13)x),$$ $j=1,\ldots,13$, and $F_{0,0}(t) = t^2$.
After generating the true functional values, the data $Y_{ijt}$ is generated from N$(a_{0,j}(X_i'\beta_0, Z_{i}'\eta_0) - a_{1,j}(X_i'\beta_0, Z_{i}'\eta_0)F_0(t),1)$.

{\it{Data generation for the linear case:}}

In this case, the true model is the one given in \eqref{linear}. All the steps for generating $X$, $Z$, $\beta_0$ and $\eta_0$ are as given above. The coefficients $\gamma_{1,j}$ and $\gamma_{2,j}$, $j=1,\ldots,13$, are generated from N$(0,1)$. After generating the design matrix, the data $Y_{ijt}$ is generated from N$(X_i\beta_0 + Z_i\eta_0 + \gamma_{1,j} - (X_i\beta_0 + Z_i\eta_0 + \gamma_{2,j})t,1)$

For the data generation, we set the error standard deviation $\sigma_0=1$. We compare the prediction MSEs across all the methods. We split the whole data into two equal parts for training and testing. We compare the performance of our method with LASSO (\cite{lasso}), SCAD (\cite{SCAD}) and horseshoe \citep{carvalho2010horseshoe} on testing dataset based estimates of parameters from the training dataset. We use the R package {\tt glmnet} for LASSO and {\tt ncvreg} for SCAD. We developed and used an HMC sampler for the horseshoe prior. For sample sizes 200, 500 and 1000, we gather the mean squared error (MSE) values both for non-linear and linear cases. We use half of the sample for training and the remaining half for testing. Among other parameters, we consider thirteen regions in total, five time points and vary the value of $p$ as 5000, 10000 and 20000. We include one case for ultra high dimension of $p=100000$ and sample size $200$. We set $M_1 = 0.1, M_2=10$ and tune $q$ to ensure a good acceptance rate and desired model size (sum of $\gamma_i$'s) across MCMC samples. We consider to maintain the desired model size between 20 and 30 at each step of the MCMC iterations. The results are summarized below for $50$ replications and $3000$ post burn samples after burn-in 1000 samples. The number of basis functions for spline is different across different sample sizes. To fit our model we vary the number of B-spline basis function as 8 for $n=200$, 11 for $n=500$, and 14 for $n=1000$. These numbers are chosen according to the strategy described in the Section~\ref{computation}. Let $I_0$ be the set of indices such that $\{i:\beta_{0i}\neq 0\}$. We have in total 26 non-zero variables. From the posterior samples of $\gamma$, we can identify the top 26 selected variables. For other methods, we select the variables with the highest 26 magnitudes in $\hat{\beta}$ and ignore the ones with very low magnitudes relative to others.  Let $\hat{X}$ denotes the final selected set of variables for different cases and let $X_{I_0}$ be the true set of variables. To examine variable selection, we report the maximum canonical correlation between $\hat{X}$ and $X_{I_0}$ for each case in the bracket.

\begin{table}[htbp]
\caption{Comparison of the proposed high-dimensional single-index model with LASSO, SCAD and horseshoe in terms of MSE and maximum canonical correlation between selected set of variables and true set of variables in the bracket for non-linear case.}
\label{nonlinearsim}
\centering
\begin{tabular}{rrrrrr}
\\
  \hline
  Total  & Dimension  & SIM  & LASSO &SCAD & Horseshoe\\
   sample size & of $\beta$ ($p$)  & MSE  &  MSE& MSE&MSE\\
  \hline
200 & 5000 & 3.33 (0.84) & 6.93 (0.37) & 7.32 (0.71) &6.24(0.60)\\ 
500 & 5000 & 3.43(0.75) & 6.11 (0.32) & 6.03 (0.69)&7.46(0.61) \\ 
1000 & 5000 & 3.27(0.76) & 6.36 (0.18) &  7.08 (0.69)&6.89(0.60)\\ 
200 & 10000 & 3.42(0.84) & 6.32 (0.12) & 7.16 (0.69) &8.33(0.59)\\ 
500 & 10000 & 3.09(0.74) & 6.69 (0.34) & 7.80 (0.66)&7.18(0.58)\\ 
1000 & 10000 & 3.25 (0.77) & 6.01 (0.35) & 7.13(0.67)&7.84(0.58)\\ 
200 & 20000 & 3.40 (0.83) & 6.88 (0.07)& 7.75 (0.70)&5.24(0.61) \\ 
500 & 20000 & 3.31 (0.74) & 6.54 (0.34) & 7.20 (0.67)&7.54(0.60) \\ 
1000 & 20000 & 3.15 (0.75) & 6.12 (0.25)& 7.03 (0.70)&7.86(0.55)\\ 
   \hline
\end{tabular}
\end{table}

\begin{table}[htbp]
\caption{Comparison of the proposed high-dimensional single-index model with LASSO, SCAD and horseshoe in terms of MSE and maximum canonical correlation between selected set of variables and true set of variables in the bracket for linear case.}
\label{linearsim}
\centering
\begin{tabular}{rrrrrr}
\\
  \hline
 Total  & Dimension  & SIM  & LASSO &SCAD&Horseshoe \\
   sample size & of $\beta$ ($p$)  & MSE  &  MSE& MSE&MSE\\
  \hline
200 & 5000 & 1.67 (0.92)& 1.02 (0.59) & 1.01 (0.86)& 1.01 (0.72)\\ 
500 & 5000 & 1.48 (0.71) & 1.01 (0.43) & 1.02 (0.74) & 1.01 (0.66)\\ 
1000 & 5000 & 1.61 (0.71) & 1.00 (0.49) & 1.02 (0.70) & 1.00 (0.66)\\ 
200 & 10000 & 1.61 (0.78) & 1.02 (0.39) & 1.01 (0.76) & 1.00 (0.78)\\ 
500 & 10000 & 1.48 (0.65) & 1.01 (0.38) & 1.02 (0.66) &1.02 (0.58)\\ 
1000 & 10000 & 1.65 (0.66) & 1.01 (0.40) & 1.02 (0.64)&1.01 (0.59)\\ 
200 & 20000 & 1.35 (0.79) & 1.03 (0.22) & 1.01 (0.73) &1.01 (0.65)\\ 
500 & 20000 & 1.24 (0.72)& 1.02 (0.35) & 1.01 (0.73) &1.00 (0.66)\\ 
1000 & 20000 & 1.31 (0.70) & 1.01(0.45) & 1.02(0.65) &1.01 (0.60)\\ 
   \hline
\end{tabular}
\end{table}

From Table~\ref{nonlinearsim}, we infer that the performance of the proposed Bayesian method based on the high-dimensional single-index model is always much better than the LASSO, the SCAD and horseshoe for the non-linear case. For the linear case in Table~\ref{linearsim}, it is competitive with linearity based methods like the LASSO, the SCAD and horseshoe. This is natural as the LASSO, the SCAD or the horseshoe use more precise modeling information which the semiparametric methods cannot use. But, in terms of maximum canonical correlations, our model outperforms in the non-linear case and remains extremely competitive for the linear case.

\section{Real-data analysis}
\label{realdata}

\subsection{Modification of the model for real data application}

\subsubsection{Incorporating random effect and region wise varying effect}
As the data are longitudinal, it is reasonable to add subject specific random effect ($\tau_i$) in the model and vary the effect of the low dimensional covariates region-wise. The new modified model will then become
\begin{align}
Y_{ijt} &= F_{ijt} + \tau_i + \varepsilon_{ijt}, \quad \varepsilon_{ijt} \sim \mathrm{N}(0, \sigma^2), \nonumber\\
F_{ijt} &= a_{0,j}(X_i'\beta, Z_i'\eta_j) - a_{1,j}(X_i'\beta, Z_i'\eta_j)F_0(t), \label{re-model}
\end{align}
$t=1,\ldots,T_i$ with $1\leq T_i\leq 6$, $j=1,\ldots,14$, $i=1,\ldots,748$.

{\it{Prior on the random effects}}

We put a Dirichlet process scale mixture of normal prior on the random effect distribution. 
\subsubsection{Region-wise varying effect with no SNP}

To compare the nonlinear model with the linear model of \cite{Hostage}, we also fit the following model without the SNPs,
\begin{align}
Y_{ijt} &= F_{ijt} + \tau_i + \varepsilon_{ijt}, \quad \varepsilon_{ijt} \sim \mathrm{N}(0, \sigma^2), \nonumber\\
F_{ijt} &= a_{0,j}(Z_i'\eta_j) - a_{1,j}(Z_i'\eta_j)F_0(t), \label{rere-model}
\end{align}
$t=1,\ldots,T_i$ with $1\leq T_i\leq 6$, $j=1,\ldots,14$, $i=1,\ldots,748$. 

\subsubsection{Corresponding Linear Model}
We compare the performance of our above non-linear models with following linear model, 
$$Y_{ijt} = H_{ijt} + \tau_i + \varepsilon_{ijt}, \quad \varepsilon_{ijt} \sim \mathrm{N}(0, \sigma^2),$$
where $t=1,\ldots,T_i$ with $1\leq T_i\leq 6$, $j=1,\ldots,14$, $i=1,\ldots,748$, and 
\begin{align}
H_{ijt} =& \varrho_{j0} + \varrho_{j, \text{M}}^0Z_{i,\text{M}}+\varrho_{j, \text{AD}}^0Z_{i,\text{AD}} +\varrho_{j, \text{NC}}^0Z_{i, \text{NC}}+\varrho_{j, \text{Allele4}}^0Z_{i, \text{Allele4}}\nonumber\\&+\varrho_{j, \text{Allele2}}^0Z_{i, \text{Allele2}}+\varrho_{j, \text{Age}}^0Z_{i,\text{Age}}+\varrho_{j, \text{AD,Allele2}}^0Z_{i,\text{AD}}Z_{i, \text{Allele2}}\nonumber\\&+\varrho_{j, \text{AD,Allele4}}^0Z_{i,\text{AD}}Z_{i, \text{Allele4}}+\varrho_{j, \text{NC,Allele2}}^0Z_{i,\text{NC}}Z_{i, \text{Allele2}}\nonumber\\&+\varrho_{j, \text{NC,Allele4}}^0Z_{i,\text{NC}}Z_{i, \text{Allele4}}-\big[\varrho_{j1} + \varrho_{j, \text{M}}^1Z_{i,\text{M}}+ \varrho_{j, \text{AD}}^1Z_{i,\text{AD}} +\varrho_{j, \text{NC}}^1Z_{i, \text{NC}}\nonumber\\&+\varrho_{j, \text{Allele4}}^1Z_{i, \text{Allele4}}+\varrho_{j, \text{Allele2}}^1Z_{i, \text{Allele2}}+\varrho_{j, \text{Age}}^1Z_{i,\text{Age}}\nonumber\\&+\varrho_{j, \text{AD,Allele2}}^1Z_{i,\text{AD}}Z_{i, \text{Allele2}}+\varrho_{j, \text{AD,Allele4}}^1Z_{i,\text{AD}}Z_{i, \text{Allele4}}\nonumber\\&+\varrho_{j, \text{NC,Allele2}}^1Z_{i,\text{NC}}Z_{i, \text{Allele2}}+\varrho_{j, \text{NC,Allele4}}^1Z_{i,\text{NC}}Z_{i, \text{Allele4}}\big]t. \label{linearmodel}
\end{align}

We have the volumetric measurement data for the total thirteen brain regions along with the summary measure of the whole brain over time for 748 individuals. For each individual, the covariate information is summarized in Table~\ref{tabledemo}. The baseline subject for our analysis is a female individual with average age and no cognitive impairment.

\begin{table}[htbp]
\caption{Demographic table to summarize data in terms of number of no cognitive impaired (NC), Alzheimer's disease (AD) and mildly cognitive impaired (MCI) participants along with the age across the two gender groups male and female.}
\centering
\begin{tabular}{rrr}
\\
  \hline
 & Female & Male \\ 
  \hline
Number of NC participants & 99 & 114 \\ 
  Number of AD participants & 84 & 94 \\ 
  Number of MCI participants & 122 & 235 \\ 
  Number of participants with APOEgene allele2&29&29\\
  Number of participants with APOEgene allele4&150&223\\
 Average age & 73.51 & 74.60 \\ 
    Standard deviation of Age  & 6.67 & 6.80 \\ 
   \hline
\end{tabular}
\label{tabledemo}
\end{table}
We first fit the model in ~\eqref{rere-model} and the following linear model in~\eqref{linearmodel} in accordance with \cite{Hostage} with same set of covariates and interactions between APOE and disease states. Then we compare the prediction MSE. 
Prediction error gives us the predictive performance and fitted relative MSE helps to judge the reliability of inference. We consider 17 basis B-spline functions for univariate and $17^2$ basis functions for bivariate cases. The estimates are based on 5000 post-burn MCMC samples after burning-in the first 1000 samples.

To compare the fitted models, we calculate the prediction error in each model. To calculate the prediction error, we divide the whole dataset into training (Tr) and testing (Te) sets. We use the stratified sampling using each subject-region pair as stratum so that training will have all the individuals that belong to the testing set. This is important for prediction with a random effect in the model. The formula for prediction error will be $|\mathrm{Te}|^{-1}\sum_{(i, j, t)\in \mathrm{Te}}(Y_{ijt}-\hat{Y}_{ijt})^2$; here $|\mathrm{Te}|$ denotes total number of elements in the test set Te. The linear model gives the prediction error 3.83 
whereas that in our non-linear model hugely improves to 0.07. The model in~\eqref{re-model} with SNPs betters the prediction error to 0.06 
which is around 14\% improvement. 

After selecting the significant genes, we calculate the Bayesian information criterion (BIC) of models leaving out one of the low-dimensional covariates every time with all the genes to compare the significance. If BIC for the model leaving out covariate A is higher than the model leaving out covariate B, then covariate A is more significant than covariate B. The table below gives an ordered list of the significance of low-dimensional covariates for different regions in Tables~\ref{nlm1}. In Table~\ref{linearesti}, we show the estimates from the linear model in~\eqref{linearmodel} for the whole brain.

\begin{table}[ht]
\footnotesize\setlength{\tabcolsep}{2.5pt}
\caption{Low dimensional covariates of the model in~\ref{rere-model} with selected genes for different brain regions in their order of significance, 1 being the most significant.}
\centering
\begin{tabular}{|rllllll|}
  \hline
 & 1 & 2 & 3 & 4 & 5 & 6 \\ 
  \hline
Total Brain & Age & APOE-allele4 & APOE-allele2 & Gender & MCI & AD \\ \hline
  Ventricles & MCI & APOE-allele4 & Gender & AD & APOE-allele2 & Age \\ \hline
  Left- &&&&&&\\
  Hippocampus & APOE-allele4 & Gender & MCI & APOE-allele2 & AD & Age \\\hline 
  Right- &&&&&&\\ Hippocampus & MCI & APOE-allele4 & APOE-allele2 & Gender & AD & Age \\ \hline
  LINFLATVEN & APOE-allele4 & Gender & AD & MCI & APOE-allele2 & Age \\ \hline
  RINFLATVEN & APOE-allele4 & AD & MCI & Gender & APOE-allele2 & Age \\ \hline
  Left Medial- &&&&&&\\ Temporal & MCI & Age & APOE-allele4 & APOE-allele2 & Gender & AD \\ \hline
  Right Medial- &&&&&&\\ Temporal & Gender & APOE-allele4 & APOE-allele2 & Age & MCI & AD \\ \hline
  Left Inferior- &&&&&&\\ Temporal & APOE-allele4 & APOE-allele2 & MCI & Age & Gender & AD \\ \hline
  Right Inferior - &&&&&&\\Temporal & APOE-allele4 & MCI & APOE-allele2 & AD & Age & Gender \\ \hline
  Left Fusiform & Gender & MCI & Age & APOE-allele4 & APOE-allele2 & AD \\ \hline
  Right Fusiform & APOE-allele4 & Age & MCI & APOE-allele2 & Gender & AD \\ \hline
  Left Entorhin & APOE-allele4 & Gender & MCI & Age & AD & APOE-\\&&&&&&allele2 \\ \hline
  Right Entorhin & APOE-allele4 & Gender & Age & MCI & AD & APOE-\\&&&&&&allele2 \\ 
   \hline
\end{tabular}
\label{nlm1}
\end{table}

We map the significant SNPs from our analysis to the corresponding genes using the R package {\tt rsnps}. We tune the parameter $\gamma$ in the model to select the 20 most significant SNPs. Among those, we could map 11 of those to some genes. The significant genes from our analysis are mentioned in Section~\ref{conclusions-discussion} along with some previous studies that found the corresponding gene significant for the AD and/or cerebral atrophy. 


\begin{table}[htbp]
\centering
\caption{Estimates of covariates for Total Brain for slope from linear model.} 
\begin{tabular}{rrrrrr}
  \hline
&Value          &      Std.Error    & DF &     t-value    &     p-value   \\
\hline
time                    & 0.013     & 0.001 &2155      & 13.208 & 0.000 \\
APOEallele4:time            & 0.000      & 0.001& 2155      & 0.047  & 0.962 \\
APOEallele2:time            & $-$0.001      & 0.002& 2155      & $-$0.249   & 0.804 \\
Gender:time           & 0.003     & 0.001 &2155      & 3.077  & 0.002 \\
MCI:time                & 0.006     & 0.001 &2155      & 4.811  & 0.000 \\
AD:time                 & 0.014     & 0.002& 2155      & 6.969  & 0.000 \\
Age:time            & $-$0.002      & 0.000 &2155      & $-$4.663   & 0.000 \\
APOEallele4:MCI:time        &0.004      & 0.002 &2155      & 2.689  & 0.007 \\
APOEallele4:AD:time         & 0.004     & 0.002 &2155      & 2.105  & 0.035 \\
APOEallele2:MCI:time        & 0.001     & 0.003 &2155      & 0.292 & 0.770          \\
APOEallele2:AD:time         & $-$0.003      & 0.006& 2155      & $-$0.530   & 0.596\\
 \hline
\end{tabular}
\label{linearesti}
\end{table}

\section{Conclusions and discussion}
\label{conclusions-discussion}
We fit a bivariate single-index model to capture the volumetric change of different cortical regions in the human brain. There are both high and low-dimensional covariates as input to the unknown functions determining initial configuration and rate of change of different regions. To tackle the high-dimensional covariate within a single-index model, we provide a new technique to assign a sparse prior in this paper and propose efficient MCMC scheme using Hamiltonian Monte Carlo sampling. Posterior consistency is demonstrated. 

In our results on the real dataset, we find that allele 4 of APOE gene is always among the top three significant covariates for almost all the cases in Table~\ref{nlm1}. The fact that allele 4 of the APOE gene is significant was established in \cite{Hostage}. They used a linear model, similar to the model in~\eqref{linearmodel}. Allele 4 is not found to be significant for the linear case in Table~\ref{linearesti} as well as for several other regions in the linear case. But it is always ranked among top three most significant covariates for the non-linear model. For this dataset, linear model is not suitable. We identify 11 significant genes. There are some previous studies that also noted the significant genes from our analysis as possible candidates for the AD and/or cerebral atrophy. The genes along with associated future study citing that gene in connection with AD and/or cerebral atrophy are mentioned here SLC6A1 (cerebellum) \citep{carvill2015mutations}, KCNIP4 \citep{himes2013integration}, ADGRL3 \citep{orsini2016behavioral}, SORBS2 \citep{zhang2016impaired,lee2014sorbs2,niceta2015mutations}, LPAR3 \citep{yung2015lysophosphatidic}, SHROOM3 \citep{dickson2015targeted,freudenberg2016differential}, SORCS3 \citep{breiderhoff2013sortilin,lane2012vps10}, NPY2R \citep{lin2010genetic,schriemer2016regulators}, CWF19L2 \citep{lin2013runs}, PALLD \citep{nho2015genome} and KCNMA1 \citep{burns2011replication,tabarki2016homozygous}. In particular, SLC6A1 has been found to affect cerebellum \citep{carvill2015mutations}; ADGRL3 affects hippocampus, the prefrontal cortex, and the striatum \citep{orsini2016behavioral}; LPAR3 affects central and peripheral nervous tissues \citep{yung2015lysophosphatidic}; SORCS3 and PALLD affect hippocampus \citep{breiderhoff2013sortilin,nho2015genome}; KCNMA1 affects olfactory bulb, cortex, basal ganglia, hippocampus, thalamus, cerebellum, vestibular nuclei, and spinal cord \citep{tabarki2016homozygous}.


We have kept the low dimensional covariates fixed with time. One interesting future direction would be to modify the model to incorporate time varying covariates as well. For example, disease status of some of the participants changed during the span of this study. Although they were very small in number and we ignored them from this study, it would be useful to consider them as well using a more elaborate model. Also, we have not included MMSE scores into our analysis. This would be interesting to study the effect of that covariate as well. Our proposed model limits us to choose a unique set of genes across all the regions. We can modify to select separate sets of genes for different parts of the brain with additional computational burden. This would give us more insights about inter dependence between genes and different parts of the brain.

An package to fit high dimensional single index model as well as also to fit linear regression with Dirichlet-Laplace and horseshoe priors using HMC is {\url{https://github.com/royarkaprava/High-Dimensional-Single-index-model}}. 

\section{Acknowledgments}

Data collection and sharing for this project was funded by the Alzheimer's Disease Neuroimaging Initiative (ADNI) (National Institutes of Health Grant U01 AG024904) and DOD ADNI (Department of Defense award number W81XWH-12-2-0012). ADNI is funded by the National Institute on Aging, the National Institute of Biomedical Imaging and Bioengineering, and through generous contributions from the following: AbbVie, Alzheimer's Association; Alzheimer's Drug Discovery Foundation; Araclon Biotech; BioClinica, Inc.; Biogen; Bristol-Myers Squibb Company; CereSpir, Inc.; Cogstate; Eisai Inc.; Elan Pharmaceuticals, Inc.; Eli Lilly and Company; EuroImmun; F. Hoffmann-La Roche Ltd and its affiliated company Genentech, Inc.; Fujirebio; GE
Healthcare; IXICO Ltd.; Janssen Alzheimer's Immunotherapy Research \& Development, LLC.; Johnson \& Johnson Pharmaceutical Research \& Development LLC.; Lumosity; Lundbeck; Merck \& Co., Inc.; Meso
Scale Diagnostics, LLC.; NeuroRx Research; Neurotrack Technologies; Novartis Pharmaceuticals Corporation; Pfizer Inc.; Piramal Imaging; Servier; Takeda Pharmaceutical Company; and Transition
Therapeutics. The Canadian Institutes of Health Research is providing funds to support ADNI clinical sites in Canada. Private sector contributions are facilitated by the Foundation for the National Institutes of Health (\url{www.fnih.org}). The grantee organization is the Northern California Institute for Research and Education,
and the study is coordinated by the Alzheimer's Therapeutic Research Institute at the University of Southern California. ADNI data are disseminated by the Laboratory for Neuro Imaging at the University of Southern California.

We are grateful to the anonymous reviewer and the editor for their valuable comments that have greatly helped to improve the manuscript.

\bibliographystyle{ba}
\bibliography{atrophy}
\end{document}